# Sudden- or Frozen-density approximation in Semi-classical extended Thomas Fermi model for Wong formula used in $^{64}$Ni+$^{64}$Ni reaction

Raj Kumar* and Raj K. Gupta

*Physics Department, Panjab University, Chandigarh-160014, INDIA.*

*rajkumarfzr@gmail.com

**Abstract**

The $\ell$-summed Wong expression is applied in the semiclassical extended Thomas Fermi (ETF) method of Skyrme energy density formalism (SEDF), using both the sudden and frozen-density approximations, to the case of $^{64}$Ni+$^{64}$Ni reaction data. We find that a very much improved fit (almost exact, point to point) to the data is obtained for the frozen-density approximation with $\ell_{max}$ ($E_{c.m.}$) varying smoothly. On the other hand, the same could not be achieved in the sudden case, still requiring a modification of the barrier at both the below- and above-barrier energies. This result means to say that the phenomenon of hindrance, observed in coupled channel calculations for $^{64}$Ni+$^{64}$Ni reaction, could be explained simply on the basis of the $\ell$-summed Wong expression, at least for the frozen-density approximation in ETF method of SEDF.

## 1. Introduction

The semiclassical extended Thomas Fermi (ETF) method of Skyrme energy density formalism (SEDF) provides a convenient basis for the calculation of nucleus-nucleus interaction potential. Here, both the kinetic energy density $\tau$ and spin-orbit density $\vec{J}$ are functions of the nucleon density $\rho_q$, $q=n,p$. For the composite system, in ETF, the densities can be added either in sudden or frozen approximation [1]. Sudden-density contains the exchange terms due to anti-symmetrization whereas the frozen-density has no such effects in it. The exchange effects arise since, for the composite system, $\tau(\rho)$ and $\vec{J}(\rho)$ are expressed as functions of the $\rho_i$ ($i$=1,2 for two nuclei), which, in turn, are the sums of their nucleon densities ($\rho_i = \rho_{in} + \rho_{ip}$), with $\rho = \rho_1 + \rho_2$. On the other hand, in frozen-density, the composite nucleus densities are simply the sums of the densities of two incoming nuclei.

In a recent paper [2], using Wong's approximate $\ell=0$ barrier-based formula [3] in ETF with sudden-densities, we were able to fit reasonably well (not exactly) the fusion-evaporation cross-sections for at least the $^{64}$Ni+$^{64}$Ni reaction with the barrier modified (lowered) by varying the half-density radius $R_0$ and surface thickness $a_0$ parameters of the two-parameter Fermi density (see Fig. 1(a), dotted line). An exactly similar fit is also obtained for the frozen-densities (also, see Fig. 1(a), dashed line), using another $R_0$ and $a_0$ parameter set. In fact, the barrier modification effects are also shown [4] to be contained in the Wong's $\ell$-summed expression, neglected in its $\ell=0$ barrier-based formula [3].

In this paper, we apply the $\ell$-summed Wong expression for the first time in the semiclassical ETF method of SEDF, using both the sudden and frozen-density approximations, to the $^{64}$Ni+$^{64}$Ni reaction data [5]. The paper is organized as follows: Sections 2 and 3 give, respectively, the details of the semi-classical ETF model and the Wong formula. The calculations and results are presented in Section 4.

## 2. The semiclassical ETF model

The interaction potential in SEDF is

$$V_N(R) = \int \{H(\rho,\tau,\vec{J}) - [H_1(\rho_1,\tau_1,\vec{J}_1) + H(\rho_2,\tau_2,\vec{J}_2)]\}d\vec{r} \qquad (1)$$

with $H$ as the Skyrme Hamiltonian density, given by

$$\begin{aligned}H(\rho,\tau,J) &= \frac{\hbar^2}{2m}\tau + \frac{1}{2}t_0\left[(1+\frac{1}{2}x_0)\rho^2 - (x_0+\frac{1}{2})(\rho_n^2+\rho_p^2)\right] + \frac{1}{12}t_3\rho^{\alpha_0}\left[\left(1+\frac{1}{2}x_3\right)\rho^2 - \left(x_3+\frac{1}{2}\right)(\rho_n^2+\rho_p^2)\right] \\ &+ \frac{1}{4}\left[t_1\left(1+\frac{1}{2}x_1\right)+t_2\left(1+\frac{1}{2}x_2\right)\right]\rho\tau - \frac{1}{4}\left[t_1\left(x_1+\frac{1}{2}\right)-t_2\left(x_2+\frac{1}{2}\right)\right](\rho_n\tau_n+\rho_p\tau_p) \\ &+ \frac{1}{16}\left[3t_1\left(1+\frac{1}{2}x_1\right)-t_2\left(1+\frac{1}{2}x_2\right)\right](\vec{\nabla}\rho)^2 - \frac{1}{16}\left[3t_1\left(x_1+\frac{1}{2}\right)-t_2\left(x_2+\frac{1}{2}\right)\right]\left[(\vec{\nabla}\rho_n)^2+(\vec{\nabla}\rho_p)^2\right] \\ &- \frac{1}{2}W_0(\rho\vec{\nabla}\cdot\vec{J}+\rho_n\vec{\nabla}\cdot\vec{J}_n+\rho_p\vec{\nabla}\cdot\vec{J}_p).\end{aligned} \qquad (2)$$

Here $\rho_q$, $\tau_q$ and $J_q$ (q=n,p) are the nucleonic, kinetic energy and spin-orbit densities, respectively. $m$ is the nucleon mass. $x_i$, $t_i$, $\alpha_0$ and $W_0$ are the Skyrme force parameters, fitted by different authors to obtain better descriptions of various ground state properties of nuclei.

In Extended Thomas-Fermi (ETF) model, the kinetic energy density $\tau(\mathbf{r})$ and spin-orbit density J($\mathbf{r}$) are functions of the nucleon density $\rho_q$, included here up to second order,

defined as

$$\tau_q^{(ETF)}(\vec{r}) = \frac{3}{5}\left(\frac{3}{2}\pi^2\right)^{2/3}\rho_q^{5/3} + \frac{1}{36}\frac{(\vec{\nabla}\rho_q)^2}{\rho_q} + \frac{1}{3}\Delta\rho_q + \frac{1}{6f_q}[(\vec{\nabla}\rho_q)(\vec{\nabla}f_q) + \rho_q\Delta f_q] \quad (3)$$

$$-\frac{1}{12}\rho_q\left(\frac{\vec{\nabla}f_q}{f_q}\right)^2 + \frac{1}{2}\left(\frac{2m}{\hbar^2}\right)^2\rho_q\left(\frac{W_0}{2}\frac{\vec{\nabla}(\rho_i+\rho_q)}{f_q}\right)^2 = \tau_{TF} + \tau_S$$

$$\vec{J}_q(\vec{r}) = -\frac{2m}{\hbar^2}\frac{1}{2}W_0\frac{\rho_q}{f_q}\vec{\nabla}(\rho_i+\rho_q), \quad q=n,p;\; i=1,2 \quad (4)$$

with the effective mass form factor

$$f_q(\vec{r}) = 1 + \frac{2m}{\hbar^2}\left[\frac{1}{4}\left\{t_1(1+\frac{x_1}{2})+t_2(1+\frac{x_2}{2})\right\}\rho_i(\vec{r}) - \frac{1}{4}\left\{t_1(x_1+\frac{1}{2})-t_2(x_2+\frac{1}{2})\right\}\rho_q(\vec{r})\right]. \quad (5)$$

For the composite system, depending on the approximation used, in:

(a) Sudden approximation,

$$\tau(\rho) = \tau(\rho_1+\rho_2) = \tau(\rho_{1n}+\rho_{2n}) + \tau(\rho_{1p}+\rho_{2p})$$
$$\vec{J}(\rho) = \vec{J}(\rho_1+\rho_2) = \vec{J}(\rho_{1n}+\rho_{2n}) + \vec{J}(\rho_{1p}+\rho_{2p})$$

and, in (b) Frozen approximation,

$$\tau(\rho) = \tau_1(\rho_1) + \tau_2(\rho_2), \qquad \vec{J}(\rho) = \vec{J}_1(\rho_1) + \vec{J}_2(\rho_2) \quad \text{with}$$

$$\tau_i(\rho_i) = \tau_{in}(\rho_{in}) + \tau_{ip}(\rho_{ip}), \qquad \vec{J}_i(\rho_i) = \vec{J}_{in}(\rho_{in}) + \vec{J}_{ip}(\rho_{ip}).$$

Introducing slab approximation, we write Eq. (1) as nuclear proximity potential [6]

$$V_N(R) = 2\pi\bar{R}\int_{s_0}^{\infty}e(s)ds = 2\pi\bar{R}\int\{H(\rho) - [H_1(\rho_1) - H_2(\rho_2)]\}dZ. \quad (6)$$

with R=R$_{01}$ ($\alpha_1$, T) + R$_{02}$ ($\alpha_2$, T) + s. Here, R$_{01}$ and R$_{02}$ are the temperature (T) dependent radii of two deformed and oriented nuclei, separated by s, with minimum s$_0$-value. $\bar{R}$ is the mean curvature radius given by $\bar{R} = \frac{R_{01}R_{02}}{R_{01}+R_{02}}$, defining the geometry of the system.

For nuclear density $\rho_i$, we use the T-dependent Fermi density distribution

$$\rho_i(Z_i) = \rho_{0i}(T)[1+\exp(\frac{Z_i - R_i(T)}{a_i(T)})]^{-1} \quad (7)$$

with $-\infty \leq Z \leq \infty$, Z$_2$=R-Z$_1$, and $\rho_{0i}(T) = \frac{3A_i}{4\pi R_{0i}^3(T)}\left[1+\frac{\pi^2 a_i^2(T)}{R_i^2(T)}\right]^{-1}$ with nucleon densities $\rho_{iq}$ further defined as $\rho_{in} = \frac{N_i}{A_i}\rho_i$, $\rho_{ip} = \frac{Z_i}{A_i}\rho_i$, and the half density radii $R_{0i}(T = 0)$ and the surface thickness parameters $a_i(T = 0)$ obtained by fitting the experimental data to the polynomials in nuclear mass $A$ (= 4–209), as [7]

$R_{0i}(T=0) = 0.90106 + 0.10957A_i - 0.0013A_i^2 + 7.71458 \times 10^{-6} A_i^3 - 1.62164 \times 10^{-8} A_i^4,$

$a_i(T=0) = 0.34175 + 0.01234A_i - 2.1864 \times 10^{-4} A_i^2 + 1.46388 \times 10^{-6} A_i^3 - 3.24263 \times 10^{-9} A_i^4.$

The temperature dependence in the above formulae are then introduced as in Ref. [8]

$R_{0i}(T) = R_{0i}(T=0)[1 + 0.0005T^2], \quad a_i(T) = a_i(T=0)[1 + 0.01T^2].$

Next, the $\ell$-dependent interaction potential V(R) is given by,

$$V(R) = V_N(R, A_i, \beta_{\lambda i}, \theta_i, T) + V_C(R, Z_i, \beta_{\lambda i}, \theta_i, T) + \frac{\hbar^2 \ell(\ell+1)}{2\mu R^2},$$

where $V_N$ is the nuclear proximity potential calculated from the ETF approach and $V_C$ is the Coulomb potential [9]. The variables $V_B^\ell$, $R_B^\ell$ and $\hbar\omega^\ell$, entering in the Wong formula [3], are calculated from this potential with effect of deformations (up to $\beta_4$) and orientations included.

## 3. The Wong formula

Wong [3] defines the fusion cross-section for two deformed and oriented nuclei lying in same plane, and colliding with center-of-mass (c.m.) energy $E_{c.m.}$, in terms of angular-momentum $\ell$ partial waves, as

$$\sigma(E_{c.m.}, \theta_i) = \frac{\pi}{k^2}\sum_{\ell=0}^{\ell_{max}}(2\ell+1)P_\ell(E_{c.m.}, \theta_i); \qquad k = \sqrt{\frac{2\mu E_{c.m.}}{\hbar^2}}, \tag{8}$$

and $\mu$ as the reduced mass. Here, $P_\ell$ is the transmission coefficient for each $\ell$ which describes the penetration of barrier $V_\ell(R, E_{c.m.}, \theta_i)$, calculated in Hill-Wheller approximation [10]. An explicit summation over $\ell$ in Eq. (8) requires the complete $\ell$-dependent potentials $V_\ell(R, E_{c.m.}, \theta_i)$, with $\ell_{max}$ to be determined empirically.

Wong [3] carried out the $\ell$-summation in Eq. (8) approximately under the conditions of using only $\ell=0$ quantities, and on replacing the summation by an integral, obtained

$$\sigma(E_{cm}, \theta_i) = \frac{R^{0\,2}_B(\theta_i)\hbar\omega_0(\theta_i)}{2E_{c.m.}}\ln\left[1+\exp\left\{\frac{2\pi}{\hbar\omega_0(\theta_i)}\left(E_{cm} - V_B^{\,0}(\theta_i)\right)\right\}\right] \tag{9}$$

which on integrating over $\theta_i$ gives $\sigma(E_{c.m.})$,

i.e. $\sigma(E_{cm}) = \int_{\theta_i=0}^{\pi/2} \sigma(E_{c.m.}, \theta_i) \sin\theta_1 d\theta_1 \sin\theta_2 d\theta_2.$  (10)

## 4. Calculations and results

Fig. 1(a), solid line, for the case of $\ell$-summed Wong expression, show a point to point fit to data for the frozen-density approximation, with $\ell_{max}(E_{c.m.})$ varying smoothly, as illustrated in Fig. 1(b), solid line. On the other hand, the same could not be achieved for the sudden-density (dash-dot line in Fig. 1(a)), still requiring a modification of the barrier

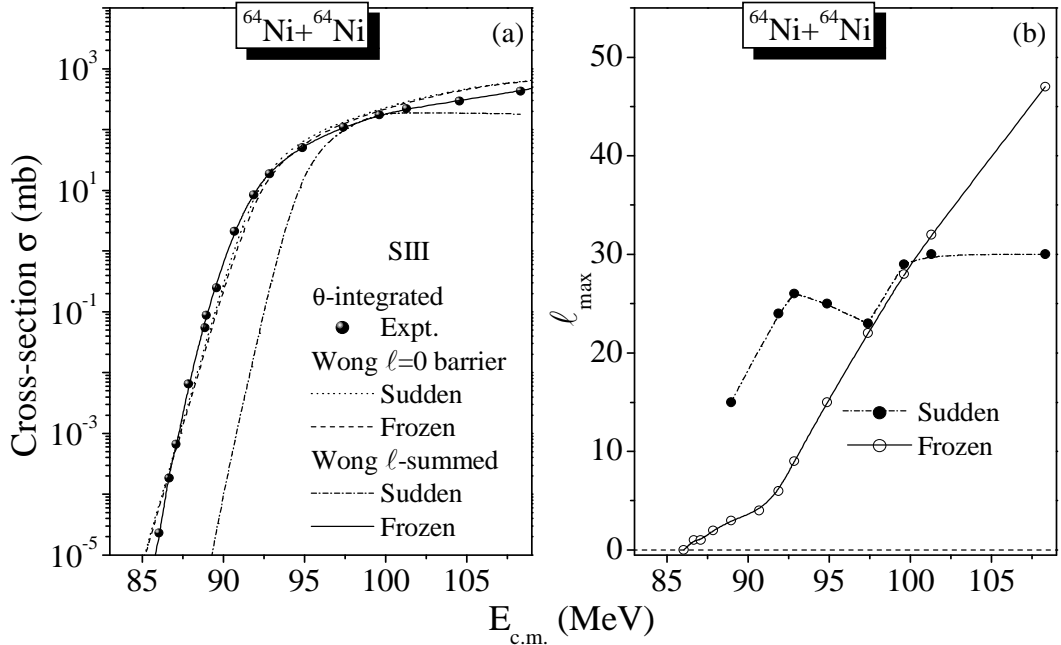

Fig. 1: (a) Cross-section fitted under sudden- and frozen-densities using $\ell=0$ barrier-based Wong formula and Wong $\ell$-summed expression for Skyrme force SIII, (b) variation of $\ell_{max}$ with $E_{c.m.}$ for the case of $\ell$-summed Wong expression.

at both the below- and above-barrier energies. Similar results are obtained for $^{58}Ni+^{58}Ni$ reaction [11], showing a similar exact fit to data with frozen-density in ETF. This means that the frozen-density gives appropriate barriers for the phenomenon of hindrance, observed in coupled channel calculations for these reactions [12], to be explained simply on the basis of $\ell$-summed Wong expression.

Concluding, the $\ell$-summed Wong expression, using the barriers calculated in frozen-density approximation in semi-classical ETF method based on SEDF, describe the $^{64}$Ni+$^{64}$Ni data on cross-sections without introducing any effects of barrier modifications.

**Acknowledgement**

The financial support from CSIR, New Delhi and Department of Science and Technology (DST), Govt. of India, is gratefully acknowledged.